\documentclass[pra,superscriptaddress,twocolumn,aps,amssymb,amsmath]{revtex4-1}

\usepackage{bm}
\usepackage{graphicx}
\usepackage{color}
\usepackage{soul} 
\usepackage{mathtools}
\usepackage{braket}
\usepackage{amsfonts}
\usepackage{textcomp} 
\usepackage{microtype} 
\usepackage{mathrsfs}
\usepackage{hyperref}
\hypersetup{colorlinks=true,  citecolor=blue}
\usepackage[all]{hypcap} 

\makeatletter
\DeclareRobustCommand{\cev}[1]{%
  {\mathpalette\do@cev{#1}}%
}
\newcommand{\do@cev}[2]{%
  \vbox{\offinterlineskip
    \sbox\z@{$\m@th#1 x$}%
    \ialign{##\cr
      \hidewidth\reflectbox{$\m@th#1\vec{}\mkern4mu$}\hidewidth\cr
      \noalign{\kern-\ht\z@}
      $\m@th#1#2$\cr
    }%
  }%
}
\makeatother 

\newcommand{\eq}[1]{\begin{align}#1\end{align}}
\newcommand{\ba}{\begin{array}}

\newcommand{\ea}{\end{array}}
\newcommand{\bit}{\begin{itemize}}
\newcommand{\eit}{\end{itemize}}

\begin{document}

\title{
Programmable directional emitter and receiver \\ of itinerant microwave photons in a waveguide}

%

\author{Nicolas Gheeraert}
\thanks{These two authors contributed equally to this work.}
\affiliation{Research Center for Advanced Science and Technology (RCAST), The University of Tokyo, Meguro-ku, Tokyo 153-8904, Japan}
\affiliation{Department of Condensed Matter Physics and Materials Science, Tata Institute of Fundamental Research, Homi Bhabha Road, Mumbai 400005, India}

\author{Shingo Kono}
\thanks{These two authors contributed equally to this work.}
\affiliation{Center for Emergent Matter Science (CEMS), RIKEN, Wako, Saitama 351-0198, Japan}

\author{Yasunobu Nakamura}
\affiliation{Research Center for Advanced Science and Technology (RCAST), The University of Tokyo, Meguro-ku, Tokyo 153-8904, Japan}
\affiliation{Center for Emergent Matter Science (CEMS), RIKEN, Wako, Saitama 351-0198, Japan}

\begin{abstract}
We theoretically demonstrate dynamically selective directional emission and absorption of a single itinerant microwave photon in a waveguide. The proposed device is an artificial molecule composed of two qubits coupled to a waveguide a quarter-wavelength apart. By using simulations based on the input--output theory, we show that upon preparing an appropriate entangled state of the two qubits, a photon is emitted directionally as a result of the destructive interference occurring either at the right or left of the qubits. Moreover, we demonstrate that this artificial molecule possesses the capability of absorbing and transmitting an incoming photon on-demand, a feature essential to the creation of a fully inter-connected one-dimensional quantum network, in which quantum information can be exchanged between any two given nodes.
\end{abstract}

\date{\today}

\maketitle

{\it Introduction}--- As the control over superconducting qubits in a single module is reaching levels near the requirements for fault-tolerant quantum computing~\cite{arute_quantum_2019}, significant efforts have been devoted to the development of more complex quantum architectures with a larger number of qubits. In this context, one strategy is to build quantum networks~\cite{kimble_quantum_2008,northup_quantum_2014,ritter_elementary_2012,cirac_quantum_1997} consisting of nodes that exchange quantum information in the form of microwave photons through their coupling to interconnecting waveguides. A key element for such a platform to function is the ability to distribute quantum information to the desired parts of the network. More specifically, in the case of one-dimensional~(1D) networks, the capability of nodes to emit their excitations in a chosen direction---\mbox{right or left}---along the waveguide, as well as to transfer them beyond their nearest neighbors, is essential~\cite{vermersch_implementation_2016,ramos_non-markovian_2016}.\\ 
\begin{figure}[t]
\includegraphics[width=0.99\linewidth]{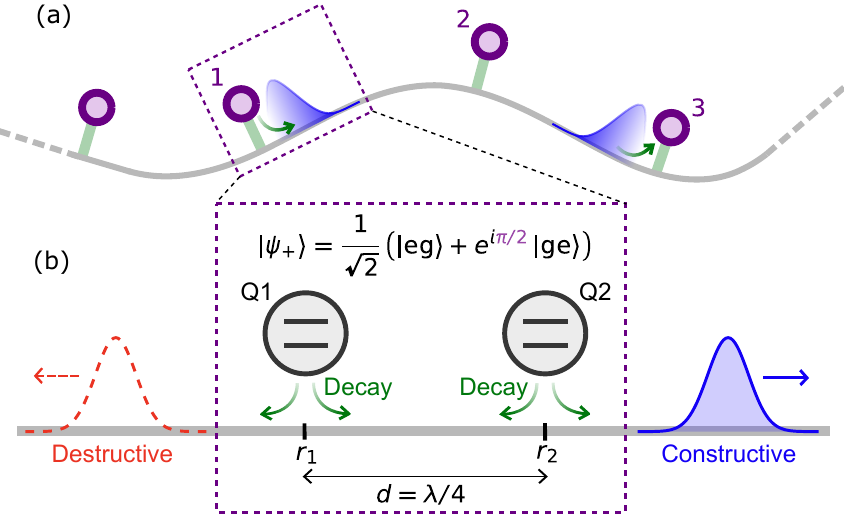} 
\caption{(a)~ 1D network of quantum nodes (purple circles) able to emit or absorb directionally in a 1D waveguide (gray line). In the illustration, node~1 is emitting a photon to the right, while node~3 is absorbing the same photon after it has been transmitted through node~2. (b)~Schematic of a programmable directional emitter/receiver which consists of two qubits coupled to the waveguide a quarter-wavelength apart. For the directional emission, the two qubits are prepared in an appropriate entangled state, the phase of which determines the direction of the emission. For a phase $\pi/2$~($-\pi/2$) the photon is emitted right~(left).
}
\label{figure1}
\end{figure}
\indent In this regard, the field of quantum optics has made significant strides forward by making use of the chiral coupling that naturally arises in photonic waveguides and nanofibers~\cite{bliokh_extraordinary_2014,bliokh_spinorbit_2015,  lodahl_chiral_2017,lodahl_quantum-dot_2018}. In these types of nano-structures, the confinement of light induces a locking of the local polarization of a photon with its direction of propagation, a phenomenon that has made possible the development of various optical devices with chirality~\cite{sayrin_nanophotonic_2015,scheucher_quantum_2016,sadgrove_polarization_2017}, including directional emitters of single photons~\cite{petersen_chiral_2014,mitsch_quantum_2014,sollner_deterministic_2015,coles_chirality_2016}.\\
\indent In quantum microwave circuits, although certain chiral effects have been studied theoretically \cite{mirza_multiqubit_2016,mirza_two-photon_2016}, such an on-demand directional emitter has not yet been realized experimentally, thus limiting the prospect of efficiently distributing quantum information in microwave 1D networks. Nonetheless, quantum state transfer was previously demonstrated between two nodes located at the two ends of a transmission line~\cite{wenner_catching_2014,kurpiers_deterministic_2018, kurpiers_quantum_2019, axline_-demand_2018, leung_deterministic_2019, campagne-ibarcq_deterministic_2018}, a configuration in which each node has only one possible direction of emission and absorption. Moreover, the coherent exchange between an artificial atom and two resonant atom-like mirrors~\cite{chang_cavity_2012} via an open waveguide was demonstrated by making use of the qubit--qubit cooperative effects provided by the coupling to the waveguide continuum of modes~\cite{mirhosseini_cavity_2019}. Extended microwave 1D networks consisting of large arrays of nodes, however, will require a new strategy for efficiently transferring the photons to the relevant nodes of the network~\cite{guimond_unidirectional_2020}. \\
\indent In this paper, we propose a quantum node consisting of an artificial molecule~\cite{frisk_kockum_designing_2014,kannan_waveguide_2020}, capable of directionally emitting and absorbing single photons along a one-dimensional waveguide. The node is composed of two qubits Q1 and Q2 coupled to a 1D waveguide a quarter-wavelength ($\lambda /4$) apart at positions $r_1$ and $r_2$, as depicted in Fig.~\ref{figure1}(b), where $\lambda$ is the microwave wavelength propagating along the waveguide at the qubit frequency. When prepared in the entangled state $\ket{\psi_\pm} =(1/\sqrt{2}) \left(  \ket{\rm eg} +e^{\pm i\pi/2}\ket{\rm ge}   \right)$, the node emits a single photon, either leftwards or rightwards depending on the sign of the phase. In contrast to the case with intrinsic chirality of photonic waveguides that is due to polarization--momentum locking, here the directionality arises from a simple interference effect~\cite{zhang_quantum_2018,zhang_heralded_2019, guimond_unidirectional_2020}. To see this, consider the case where the two qubits are prepared with a phase $+\pi/2$. Due to the delocalized nature of the quantum state, the single excitation stored in the device will be simultaneously emitted from both qubits, with a $\pi/2$ phase difference between the radiation generated by Q1 and Q2. Upon traveling the $\lambda/4$ distance to Q1, the leftward emission of Q2 further accumulates a $\pi/2$ phase difference, bringing it exactly out of phase with the emission from Q1. This results in the destructive interference of the leftward propagating photon (dashed red wave packet). Similarly, one can easily see that the photon amplitudes moving rightwards interfere constructively (blue wave packet). It is worth noting that this argument is also valid for cases with time-dependent relaxation rates as long as the time scale is much longer than that of the propagation between the two qubits. Thus, by modulating the relaxation rates, we can control the temporal mode shape of the emitted photons. Importantly, this proposal can be assembled using 2D circuit components commonly used in circuit QED experiments. A concrete example of how this could be done is provided in Appendix \ref{appendixG}.
\\
\indent This two-qubit molecule can be used to create a fully inter-connected 1D quantum network, where any two nodes can exchange quantum information with one another. In the example shown in Fig.~\ref{figure1}(a), a photon emitted rightward by node 1 is reabsorbed by node 3, after having been transmitted through node 2. After briefly outlining the model based on the input--output theory, we demonstrate the capability of an individual node to perform these three essential tasks required for a fully inter-connected 1D network ---selective directional emission, absorption, and transmission of a photon. Although we focus here on the exchange of a single photon, these protocols can be extended to the manipulation of any quantum information stored, for example, in the form of a superposition of the single-photon and vacuum states. \\
\indent {\it The model---}
The two qubits of the aforementioned artificial molecule present two cooperative effects mediated by virtual photons in the waveguide: an energy-exchange interaction and a correlated decay~\cite{van_loo_photon-mediated_2013,lalumiere_input-output_2013}. For an inter-qubit distance of $d=\lambda/4$, it turns out that the energy-exchange interaction is maximized while the correlated decay vanishes. The master equation of the artificial molecule coupled to the waveguide is therefore given by
\eq{
\frac{d\hat{\rho}}{dt}= &-i\left[\hat{H}_{\rm s}+\hat{H}_J+\hat{H}_{\rm c},\:\hat{\rho}\right]
+\gamma(t) \sum_{j=1,2}  D(\hat L_j)\hat \rho,
\label{master_equation}
}
\noindent where $\hat{L}_j$ is the lowering operator of each qubit and $D(\hat A)=\hat A\hat{\rho}\hat A^\dag-\frac{1}{2}\{\hat A^\dag \hat A,\hat{\rho}\}_+$ denotes the Lindblad super-operator. The time-varying external coupling rate $\gamma(t)$ of both qubits can be implemented by using a SQUID~\cite{yin_catch_2013}. The system Hamiltonian of two qubits with the excitation frequency of $\omega_\mathrm{p}$ is given by 
\eq{
\hat{H}_{\rm s}=\sum_{j=1,2} \omega_{\rm p}\hat{L}_j^\dag\hat{L}_j.
}
In addition, the effective Hamiltonian describing the qubit--qubit interaction mediated by the waveguide is given by 
\eq{
\hat{H}_J = J(t)\left( \hat L_1^\dag \hat L_2 + \hat L_2^\dag\hat L_1\right),
}
where 
\eq{
J(t)= \gamma(t)/2
} 
is the coupling strength for $d=\lambda/4$. As we will show, the presence of this effective energy-exchange interaction, whose strength is of the same order of magnitude as the radiative decay rates, leads to a population transfer between the qubits during the process of emission (absorption), thus affecting the directionality which relies on the synchronous emission (absorption) of the qubits at all times. Therefore, an additional fast and tunable coupling~\cite{bialczak_fast_2011,chen_qubit_2014} between the qubits is required in order to implement a cancellation interaction 
\eq{
\hat{H}_{\rm c} = g_{\rm c}(t)\left( \hat L_1^\dag \hat L_2 + \hat L_2^\dag\hat L_1\right),
}
where the coupling strength is given by $g_{\rm c}(t)=-J(t)$ [Fig.~\ref{figure2}(a)]. \\
\begin{figure*}[]
\includegraphics[width=0.99\linewidth]{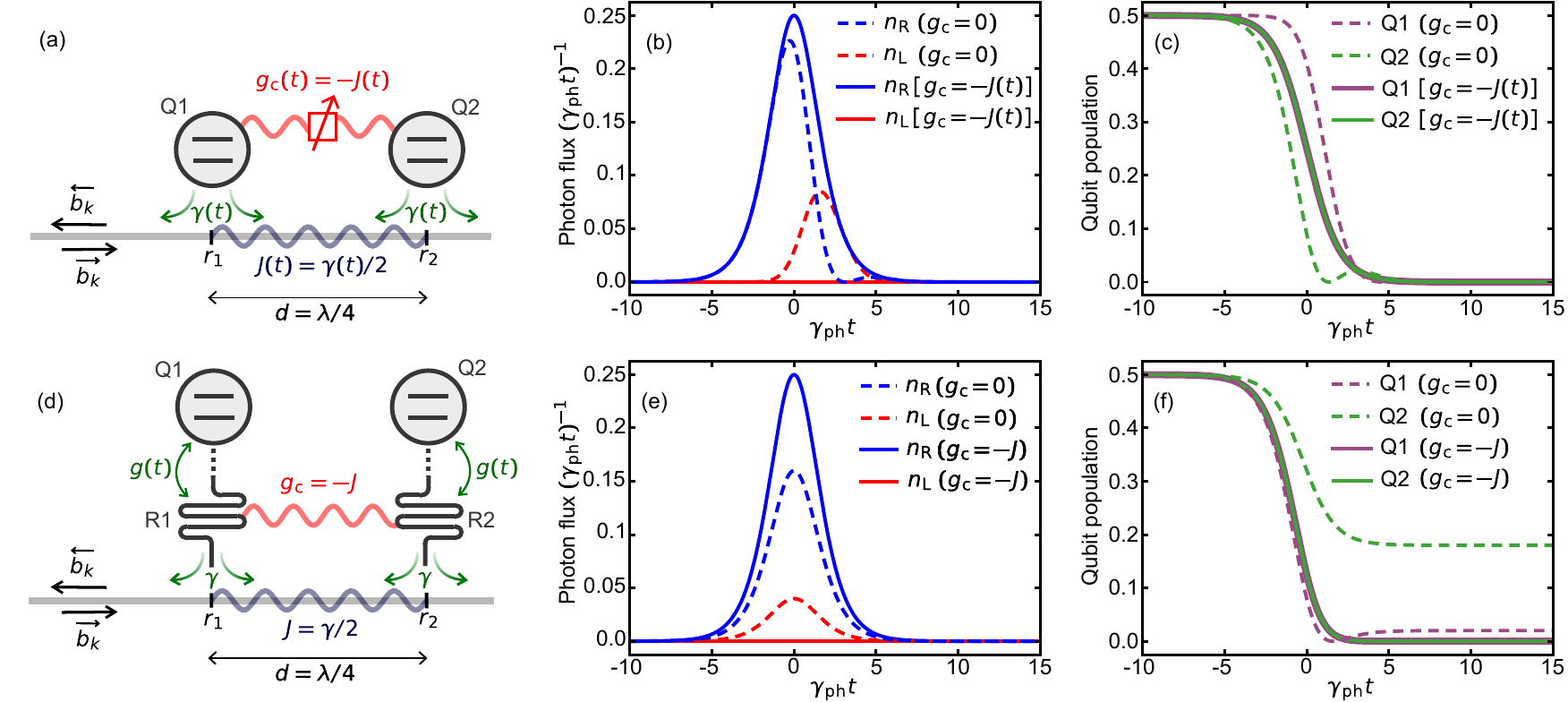}
\caption{Directional emission of shaped photons. (a)~Schematic of an artificial molecule where two qubits are directly coupled to the waveguide with a time-dependent relaxation rate $\gamma(t)$.  To cancel the waveguide-mediated interaction $J(t)$, an additional tunable coupling between the qubits is added (depicted in red). (b)~Emitted photon fluxes, $n_\mathrm{R}$ and $n_\mathrm{L}$, as a function of time normalized with the bandwidth of the photon envelope. (c)~Population of the qubit excited state as a function of time. (d)~Schematic of an extended artificial molecule where two qubits are coupled to the waveguide via resonators that decay at a constant rate. The waveguide-mediated interaction is canceled by adding a constant coupling $g_{\rm c}=-J$ between the two resonators. The plots~(e) and (f) are similar to (b) and (c), but for the setup in (d). In the four plots dashed lines represent the dynamics when the canceling interaction is turned off ($g_{\rm c}=0$).} 
\label{figure2}
\end{figure*}
\indent  To verify the directional emission, we characterize the photon fluxes emitted from the artificial molecule.
The input--output relations for the right- and left-propagating modes in the Heisenberg picture are given respectively by
\begin{equation}
\begin{aligned}
\label{boutin}
\vec{b}_{\rm out} &=  \vec{b}_{\rm in} -i\sqrt{\frac{\gamma(t)}{2}}\left(\hat{L}_1e^{+i\pi/4}-i\hat{L}_2 e^{-i\pi/4}\right),\\
\cev{b}_{\rm out} &=  \cev{b}_{\rm in} -i\sqrt{\frac{\gamma(t)}{2}}\left(\hat{L}_1e^{-i\pi/4}-i\hat{L}_2 e^{+i\pi/4}\right).
\end{aligned}
\end{equation}
Here, $\vec{b}_{\rm out}(t)$ and $\cev{b}_{\rm out}(t)$ are the outgoing field operators, and $\vec{b}_{\rm in}(t)$ and $\cev{b}_{\rm in}(t)$ are the incoming field operators at time $t$. 
Assuming that both incoming fields are in the vacuum state, the photon fluxes emitted rightwards and leftwards are obtained using these relations and the expectation values of the qubit operators:
\eq{
n_{\rm R}(t) &= \left\langle\vec{b}_{\rm out}^{\: \dag}(t)\vec{b}_{\rm out}(t)\right\rangle  \\ 
&= \frac{\gamma}{2}\left(\left\langle\hat{L}_1^\dag\hat{L}_1 \right\rangle + \left\langle \hat{L}_2^\dag\hat{L}_2 \right\rangle -i\left\langle \hat{L}_1^\dag\hat{L}_2 \right\rangle +i\left\langle \hat{L}_2^\dag\hat{L}_1 \right\rangle \right). \notag 
} 
We note that a different calculation based on input-output theory previously demonstrated this interference effect leading to directionality \cite{zhang_quantum_2018,zhang_heralded_2019}.
\\
\indent {\it Directional emission dynamics}---
We consider the emission dynamics in the molecule when it is prepared in the state 
\eq{
\ket{\psi_+} =\frac{1}{\sqrt{2}} \left(  \ket{\rm eg} +e^{i\pi/2}\ket{\rm ge} \right),
}
and when the modulation of the radiative decay rate $\gamma(t)$ is chosen so as to generate a time-symmetric photon wave packet with amplitude envelope 
\eq{
\phi(t)=\frac{1}{2}\sqrt{\gamma_{\rm ph}}\,{\rm sech}\left(\frac{\gamma_{\rm ph}t}{2}\right),
} corresponding to a photon flux $|\phi(t)|^2$ (see Appendix~\ref{appendixF} for details). Note that in the rest of the paper, all the prepared itinerant photons are assumed to have this specific shape. In Fig.~\ref{figure2}(b), we show the emitted photon flux as a function of time, with dashed lines corresponding to the absence of cancellation of the waveguide-mediated coupling $[g_c(t)=0]$, and with solid lines to perfect cancellation $[g_c(t)=-J(t)]$.  Clearly, with no cancellation, a significant part of the photon flux is still emitted leftward, a behavior that is easily understood by observing the dynamics of the qubit populations shown in Fig.~\ref{figure2}(c): as the decay rate $\gamma (t)$ is turned on, the population from qubit Q2 is substantially transferred to qubit Q1 due to the coupling $J(t)$. This population imbalance causes an imperfect interference of the radiation coming from the two qubits, resulting in a strong reduction in directionality.\\
\indent It is interesting to note that the waveguide-mediated interaction is directly proportional to the radiative decay rate, implying that with a constant decay rate to the waveguide, the cancellation coupling would also remain constant, thus making the experimental implementation significantly simpler. This is the idea behind the second design presented in Fig.~\ref{figure2}(d), where two identical resonators, R1 and R2, now mediate the transfer of the photons emitted by the qubits to the waveguide. Moreover, to allow on-demand photon generation and its pulse shaping, the qubits need to be coupled to their respective resonators with tunable energy-exchange interactions. For instance, such interactions can be experimentally realized by using a microwave-induced parametric coupling~\cite{pechal_microwave-controlled_2014} or a tunable SQUID-inductance-mediated coupling~\cite{chen_qubit_2014}. 
The constant radiative decay of the resonators $\gamma$ and the tunable qubit--resonator coupling $g(t)$ induce the effectively tunable qubit decay to the waveguide, making it possible to realize pulse shaping of a directionally emitted photon. 
In other words, the system Hamiltonian in the master equation [Eq.~(\ref{master_equation})] is replaced with
\eq{
\hat{H}_{\rm s} = \sum_{j=1,2} \left[\omega_{\rm p}\hat{L}_j^\dag \hat{L}_j + \omega_{\rm p}\hat{\sigma}_j^\dag \hat{\sigma}_j  + g(t)(\hat{L}_j^\dag\hat{\sigma}_j+\hat{L}_j\hat{\sigma}_j^\dag)\right],
}
where $\hat{L}_j$ is the lowering operator of resonator R$j$ and $\hat{\sigma}_j$ is the lowering operator of qubit Q$j$.
Conveniently, in this configuration the waveguide-mediated interaction, described by $\hat{H}_J$, couples the transfer resonators with a constant value $J =\gamma/2$, hence requiring no more than a constant coupling $g_{\rm c} =-\gamma/2$ in $\hat{H}_{\rm c}$ to be applied between the resonators. \\
\indent In Figs.~\ref{figure2}(e) and \ref{figure2}(f), we compare the emitted photon fluxes from the now enlarged artificial molecule, with and without coupling cancellation. As expected, the presence of the constant coupling between the resonators allows the two qubits to decay in synchrony, leading to perfect destructive interference on one side of the waveguide. One could expect, however, that fabrication imprecisions could lead to around 10\% error in the strength of the built-in cancellation coupling $g_{\rm c}$, resulting in a residual waveguide-mediated interaction. Remarkably though, we find that a residual coupling of 10\% only leads to a reduction of about 0.5\% of the directionality~(see Appendix~\ref{appendixH}), thus demonstrating the robustness of directionality against residual waveguide-mediated couplings. 
Hereinafter, we focus on the more feasible design involving an artificial molecule having transfer resonators,
although we have confirmed that the case without transfer resonators also functions in the same way.\\
\indent {\it Absorption of an itinerant photon}---
\begin{figure}[t]
\includegraphics[width=0.99\linewidth]{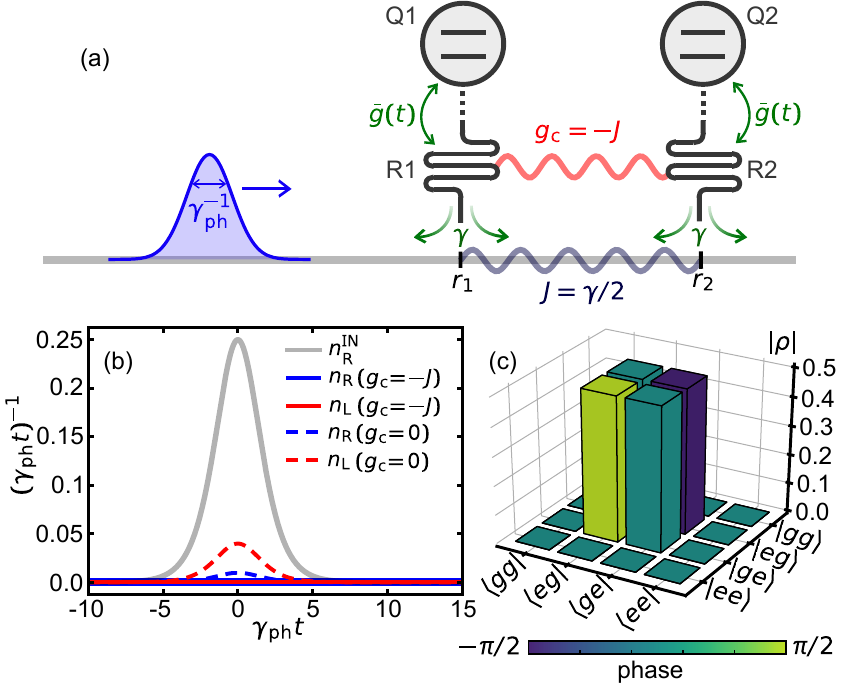}
\caption{Directional absorption. (a) Illustration of the protocol: to absorb an itinerant photon, the modulation of the qubit--resonator coupling $\bar{g}(t)$ should be the time-reversal $g(-t)$ of the modulation $g(t)$ that is required to generate the photon. (b)~Scattered photon flux. The gray line depicts the input photon flux $n_\mathrm{R}^\mathrm{IN}$. 
(c) Reduced density matrix of the final state of the artificial molecule for $g_{\rm c}=-J$.}
\label{figure3}
\end{figure}
\begin{figure}[h]
\includegraphics[width=0.99\linewidth]{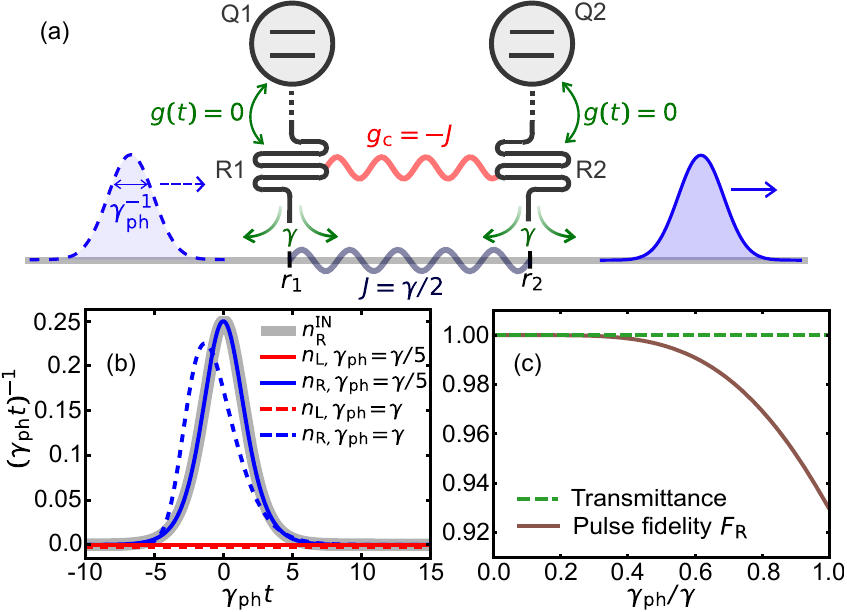}
\caption{Transmission of a photon across an artificial molecule when the qubits are decoupled. (a) Illustration of the node in `pass-through' mode: the qubit--resonator coupling $g(t)$ is switched off. (b) Transmitted and reflected photon flux for two different values of the photon wave packet bandwidth~$\gamma_\mathrm{ph}$. Note that the time taken by the photon to be transmitted is subtracted from the scattered fluxes so that the input and scattered flux curves can be directly compared. (c) Transmitted pulse fidelity~$F_\mathrm{R}$~(solid line) and transmittance~(dashed line) versus the bandwidth~$\gamma_\mathrm{ph}$ of the photon wave packet. The bandwidth is normalized by the coupling rate of the transfer resonator, $\gamma$. }
\label{figure4}
\end{figure}
Just as a usual single superconducting qubit cannot be made to emit directionally, so can it not fully absorb a single photon coming from a single direction. An artificial molecule is therefore also required at the receiving node. The absorption protocol is depicted in Fig.~\ref{figure3}(a), where the same itinerant photon that was emitted rightwards in the previous paragraph is now impinging from the left. For time-symmetric wave packets, this protocol is simply a matter of time-reversing the modulation of the coupling amplitudes \cite{kurpiers_deterministic_2018}, that is to say: $g(t) \rightarrow g(-t)$. \\
\indent To numerically simulate the single-photon input to the molecule system, an ancillary qubit is coupled to the right-propagating mode as a single-photon source~(see Appendix~\ref{appendixD}). In Fig.~\ref{figure3}(b), we show the scattered photon fluxes $n_{\rm R}(t)$ and $n_{\rm L}(t)$, right and left of the device, for the case with cancellation of the effective coupling (solid lines) and without (dashed lines).  Clearly, the absorption is imperfect without cancellation, as evidenced by the significant scattered photon flux as compared to the input photon flux (thick gray line). With coupling cancellation, however, we observe no scattered flux, indicating perfect absorption of the photon. 
Moreover, in the case where the cancellation coupling is present ($g_{\rm c}=-J$), we can confirm from the final density matrix of the qubits [Fig.~\ref{figure3}(c)] that we recover the entangled state $\ket{\psi_{+}}=(1/\sqrt{2}) \left( \ket{\rm eg} +e^{+ i\pi/2}\ket{\rm ge} \right)$ that led to the rightward directional emission of a photon in the previous paragraph. \\
\indent  {\it Photon transmission}---
Finally, in order to demonstrate transmission, we consider the same situation as for the absorption protocol, except that this time we keep the qubits decoupled from the resonators at all times, i.e.~$g(t)=0$, as indicated in Fig.~\ref{figure4}(a). Because the resonators remain coupled to the waveguide, the perfect transmission of the photon through the artificial molecule is non-trivial. Figure~4(b), shows the reflected and transmitted photon fluxes, $n_{\rm L}$ and $n_{\rm R}$, for two different bandwidths of the photon wave packet, $\gamma_{\rm ph}$. Quite remarkably, the photon is transmitted through the device for both $\gamma_\mathrm{ph}$, as indicated by the absence of  left-scattered photons~(red curves). This effect is the consequence of a destructive interference between the photon amplitude reflected on resonator~R1, and the one transmitted through resonator~R1 but reflected on resonator~R2, which accumulates an additional phase $\pi$ from traveling twice the distance $\lambda/4$. The dashed green curve in Fig.~\ref{figure4}(c) confirms that the photon is fully transmitted for any value of the bandwidth of the incoming photon wave packet as long as $\gamma_\mathrm{ph} \ll \gamma$. \\
\indent Furthermore, it is found that the transmission of the photon across the two resonators involves a constant time delay of $4\gamma^{-1}$ (see Appendix~\ref{appendixE}). The transmitted fluxes $n_{\rm R}$ are hence translated back in time by this time delay in order to compare the shape of the input (gray) and transmitted (blue) photons, as displayed in Fig.~\ref{figure4}(b). In particular, this highlights the fact that the photon is significantly distorted when its bandwidth parameter $\gamma_{\rm ph}$ is chosen equal to the transfer resonator radiative decay rate $\gamma$. To quantify this distortion, the pulse fidelity $F_{\rm R}$ is computed as
\begin{equation}
\label{mode}
F_{\rm R} =  \left|\int_{-\infty}^\infty dt\:{\phi}_{\rm R}^{{\rm IN}*}(t)\phi_{\rm R}(t+4\gamma^{-1})\right|^2,
\end{equation}
where $\phi_{\rm R}^{\rm IN}(t)$ is the mode function of the input wave packet. This fidelity is plotted versus the bandwidth of the incoming photon wave packet in Fig.~\ref{figure4}(c). Interestingly, even though the distortion reaches a significant $7\%$ for $\gamma_\mathrm{ph} = \gamma$, it remains minimal for smaller photon bandwidths, giving $F_{\rm R}>99\%$ for $\gamma_{\rm ph} \lesssim 0.5 \gamma$. \\
\indent In conclusion, we have proposed a design of an artificial molecule capable of directionally emitting photons in a waveguide on demand. It was shown that although the waveguide-mediated qubit--qubit coupling inside the artificial molecule strongly perturbs the directionality of the emission, this coupling can be effectively canceled out by engineering an opposite coupling between the two emitters. 
While a dynamical cancellation is needed for a system of two qubits tunably coupled to a \mbox{waveguide}, inserting transfer resonators between the qubits and the waveguide makes it possible to realize a programmable directional emitter and receiver by using no more than a constant cancellation between the resonators. Moreover, it was demonstrated that the same artificial molecule can absorb a photon by simply implementing the reverse modulation of the qubit--resonator coupling. Finally, it was shown that this device can passively transmit incoming photons when keeping the qubits decoupled from the resonators. These three features, interchangeable on demand,  demonstrate that a one-dimensional quantum network constituted of such artificial molecules presents full interconnectivity, allowing any two given nodes of the network to exchange photons, and hence, by extension, any quantum information encoded in the nodes. Our scheme can be straightforwardly extended to multi-photon states by replacing the qubits with storage resonators which can be prepared in an appropriate superposition state using, for instance, a tunable energy-exchange interaction.
\begin{acknowledgments}
We acknowledge Pierre-Olivier Guimond and Peter Zoller for useful discussions and sharing their work~(Ref.~\cite{guimond_unidirectional_2020}) before publication. We are also grateful to Jesper Ilves and Koji Usami for discussions. We also acknowledge support from JSPS KAKENHI (18F18799), JST ERATO (JPMJER1601), and MEXT Q-LEAP (JPMXS0118068682).
\end{acknowledgments}

\appendix

\section{System for directional coupling}
\label{appendixA}
As shown in Fig.~\ref{theoretical_models}(a), we first consider an artificial molecule consisting of two qubits that are coupled to a waveguide in order to realize directional emission and absorption of single photons.
A tunable coupling between the qubit and the waveguide is required so that the emitted photon envelope can be shaped. 
In this case, setting $\hbar=1$, the system Hamiltonian is given by
\begin{equation}
\hat{H}_{\rm s} = \sum_{j=1,2} \omega_j\hat{L}_j^\dag\hat{L}_j,
\end{equation}
where  $\omega_j$ and $\hat{L}_j$ is the resonance frequency and lowering operator of qubit Q$j$, respectively.

As discussed in the main text, we also consider an artificial molecule composed of qubits and transfer resonators.
When a transfer resonator R$j$ is inserted between qubit Q$j$ and the waveguide as shown in Fig.~\ref{theoretical_models}(b), the system Hamiltonian becomes
\begin{equation}
\hat{H}_{\rm s} = \sum_{j=1,2} \left[\omega_j\hat{L}_j^\dag \hat{L}_j + \omega_j\hat{\sigma}_j^\dag \hat{\sigma}_j  + g_j(\hat{L}_j^\dag\hat{\sigma}_j+\hat{L}_j\hat{\sigma}_j^\dag)\right].
\end{equation}
In this case, $\hat{L}_j$ is the lowering operator of the transfer resonator R$j$, $\hat{\sigma}_j$ is the lowering operator of qubit Q$j$, and $g_j$ is the tunable coupling strength between resonator R$j$ and qubit Q$j$.
Both the resonance frequencies of resonator R$j$ and qubit Q$j$ are set to $\omega_j$.
The effective external coupling rates of the qubits to the waveguide via the resonators are tuned through their variable coupling strengths to the resonators that are coupled to the waveguide with a constant rate. 

\section{Cooperative effects mediated by waveguide}
\label{appendixB}
\begin{figure}[t]
\label{theoretical_models}
\begin{center}
  \includegraphics[width=80mm]{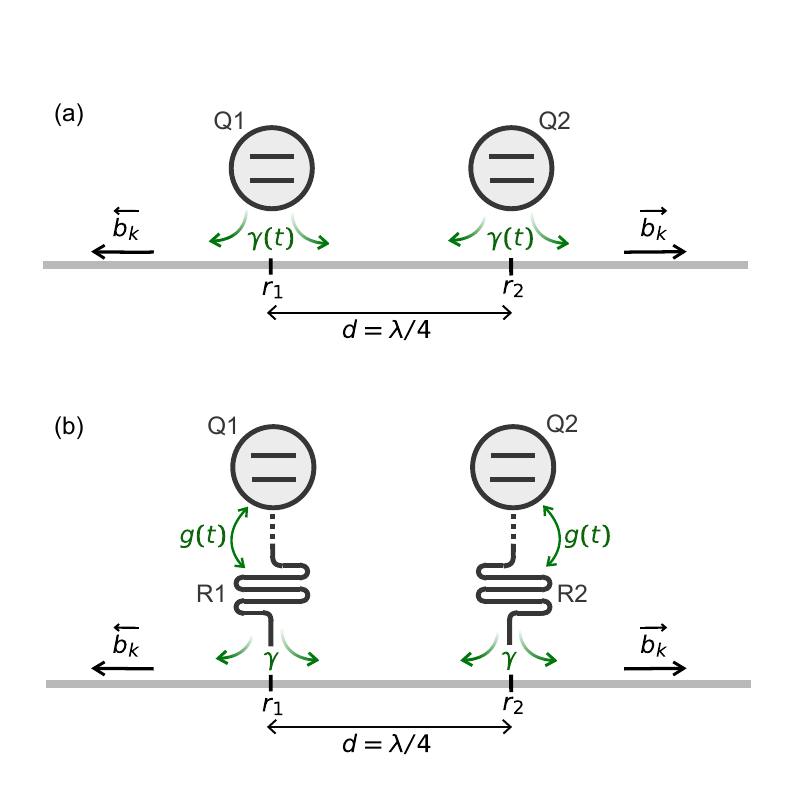} 
\caption{
Theoretical models.
(a)~Two qubits coupled to a waveguide.
(b)~Two qubit--resonator units coupled to a waveguide.
}
\end{center}
\end{figure}

When the two elements composing an artificial molecule, either qubits in the model of  Fig.~\ref{theoretical_models}(a) or transfer resonators in the model of Fig.~\ref{theoretical_models}(b), are over-coupled to a single waveguide, there are cooperative effects between the elements, mediated by virtual photons in the waveguide.
Here, we derive these cooperative effects based on the input--output relations.

We consider the setup where element E$j$ with the lowering operator $\hat{L}_j$ is coupled to the waveguide at position $r_1=-d/2$ ($r_2=d/2$) with an external coupling rate of $\gamma_1$ ($\gamma_2$), where $d$ is the distance between the two elements.
The external coupling rate is considered to be time-dependent.
Note that the external coupling rate is defined as the total radiative decay rate, counting both directions of the emission of the element coupled to the waveguide individually.
The total Hamiltonian is composed of terms corresponding to the artificial molecule ($\hat{H}_{\rm s}$), the waveguide ($\hat{H}_{\rm w}$), and the interaction between each element and the waveguide ($\hat{H}_{\rm i}$).
Setting the velocity of microwaves to unity, the waveguide Hamiltonian is given by
\begin{equation}
\hat{H}_{\rm w} = \int_{-\infty}^{\infty} dk\: |k|\hat{b}_k^\dag\hat{b}_k,
\end{equation}
where $k$ is the wavenumber of the waveguide and $\hat{b}_k$ is the annihilation operator of the mode with $k$.
The annihilation operator satisfies the commutation relation of $[\hat{b}_k,\hat{b}_{k'}^\dag]=\delta(k-k')$, where $\delta(k)$ is the Dirac delta function.
The interaction Hamiltonian under the rotating wave approximation is given by
\begin{equation}
\hat{H}_{\rm i} = \sum_{j=1,2} \int_{-\infty}^{\infty} \frac{dk}{\sqrt{2\pi}}\: \sqrt{\frac{\gamma_j}{2}} \left(\hat{L}_j^\dag  \hat{b}_k e^{+ikr_j} + \hat{L}_j\hat{b}_k^\dag e^{-ikr_j}\right),
\end{equation}
where we assume that the coupling strength is independent of the wavenumber. 

By using positive wavenumbers only, the right- and left-propagating-mode annihilation operators are relabeled as $\vec{b}_k=\hat{b}_k$ if $k>0$ and $\cev{b}_k=\hat{b}_{-k}$ if $k<0$, respectively. 
They follow $\left[\vec{b}_k,\vec{b}_{k'}^{\: \dag}\right] = \left[\cev{b}_k,\cev{b}_{k'}^{\: \dag}\right] = \delta(k-k')$.
Then, the terms related to the waveguide are rewritten as 
\begin{equation}
\hat{H}_{\rm w} = \int_{0}^{\infty} dk\: k\:\vec{b}_k^{\: \dag}\vec{b}_k +\int_{0}^{\infty} dk\: k\:\cev{b}_k^{\: \dag}\cev{b}_k
\end{equation}
and
\eq{
\hat{H}_{\rm i} = &\sum_{j=1,2} \int_{0}^{\infty} \frac{dk}{\sqrt{2\pi}}\: \sqrt{\frac{\gamma_j}{2}} \Big(\hat{L}_j^\dag  \vec{b}_k e^{+ikr_j} + \hat{L}_j\vec{b}_k^{\: \dag} e^{-ikr_j} \notag \\
&+\hat{L}_j^\dag  \cev{b}_k e^{-ikr_j} + \hat{L}_j\cev{b}_k^{\: \dag} e^{+ikr_j}\Big).
}
Since each element approximately interacts only with the propagating mode around the system frequency, it is justified to add an ancillary propagating mode with a negative frequency.
In other words, the lower limit of the integration with respect to the wavenumber can be extended to $-\infty$, i.e.\ 
\begin{equation}
\hat{H}_{\rm w} = \int_{-\infty}^{\infty} dk\: k\:\vec{b}_k^{\: \dag}\vec{b}_k +\int_{-\infty}^{\infty} dk\: k\:\cev{b}_k^{\: \dag}\cev{b}_k  
\end{equation}
and
\eq{
\hat{H}_{\rm i} = &\sum_{j=1,2} \int_{-\infty}^{\infty} \frac{dk}{\sqrt{2\pi}}\: \sqrt{\frac{\gamma_j}{2}} \Big(\hat{L}_j^\dag  \vec{b}_k e^{+ikr_j} + \hat{L}_j\vec{b}_k^{\: \dag} e^{-ikr_j} \notag \\
&+\hat{L}_j^\dag  \cev{b}_k e^{-ikr_j} + \hat{L}_j\cev{b}_k^{\: \dag} e^{+ikr_j}\Big).
}

Using the propagating-mode annihilation operators in the wavenumber space, the right- and left-propagating-mode annihilation operators in the real space are defined as
\begin{equation}
\vec{b}_r =  \int_{-\infty}^{\infty} \frac{dk}{\sqrt{2\pi}}\:\vec{b}_ke^{+ikr},\quad
\cev{b}_r =  \int_{-\infty}^{\infty} \frac{dk}{\sqrt{2\pi}}\:\cev{b}_ke^{-ikr} .
\end{equation}
The annihilation operators in the wavenumber and real spaces are connected via the Fourier transforms.
They follow the commutation relations of $\left[\vec{b}_r,\vec{b}_{r'}^{\: \dag}\right] = \left[\cev{b}_r,\cev{b}_{r'}^{\: \dag}\right] = \delta(r-r')$.

Using the total Hamiltonian, the time evolution of an arbitrary operator $\hat{O}$ supported by the artificial-molecule subspace is described in the Heisenberg picture by
\begin{widetext}
\eq{
\frac{d\hat{O}}{dt} =&\  i \left[\hat{H}_{\rm s},\hat{O}\right]
+\sum_{j=1,2} \int_{-\infty}^{\infty} \frac{dk}{\sqrt{2\pi}}\: \sqrt{\frac{\gamma_j}{2}}\bigg(\left[\hat{L}_j^\dag ,\hat{O}\right]\vec{b}_ke^{+ikr_j}+\vec{b}_k^{\: \dag} e^{-ikr_j}\left[L_j,\hat{O}\right]
+\left[\hat{L}_j^\dag ,\hat{O}\right]\cev{b}_ke^{-ikr_j}\notag +\cev{b}_k^{\: \dag} e^{+ikr_j}\left[\hat{L}_j,\hat{O}\right]\bigg). \notag \\
=&\ i \left[\hat{H}_{\rm s},\hat{O}\right]
+\sum_{j=1,2} \sqrt{\frac{\gamma_j}{2}}\bigg(\left[\hat{L}_j^\dag ,\hat{O}\right]\vec{b}_{r_j}+\vec{b}_{r_j}^{\: \dag}\left[\hat{L}_j,\hat{O}\right]  +\left[\hat{L}_j^\dag ,\hat{O}\right]\cev{b}_{r_j}+\cev{b}_{r_j}^{\: \dag}\left[\hat{L}_j,\hat{O}\right]\bigg).
\label{dOdt}
}
\end{widetext}
For simplicity, we hereinafter omit the representation of time $t$, e.g.\ $\hat{O} =\hat{O}(t)$, $\hat{L}_j =\hat{L}_j(t)$, and $\vec{b}_{r_j}=\vec{b}_{r_j}(t)$, except when we need to specify a certain time.

On the other hand, the time evolution of the annihilation operator of a right-propagating mode with the wavenumber $k$ is 
\begin{equation}
\frac{d\vec{b}_k}{dt} =  -ik\vec{b}_k-\frac{i}{\sqrt{2\pi}}\sum_{j=1,2}\sqrt{\frac{\gamma_j}{2}}\:\hat{L}_j e^{-ikr_j},
\end{equation}
which can be formally solved as
\eq{
\label{rak}
\vec{b}_k =&\ \vec{b}_k(t_{\rm i}) e^{-ik(t-t_{\rm i})} \notag \\
&-\frac{i}{\sqrt{2\pi}}\sum_{j=1,2}\int_{t_{\rm i}}^{t}dt'\:\sqrt{\frac{\gamma_j(t')}{2}}\hat{L}_j(t') e^{-ik(r_j-t'+t)},
}
where $\vec{b}_k(t_{\rm i})$ is the annihilation operator at the initial time $t=t_{\rm i}$, and $\gamma_j(t')$ and $\hat{L}_j(t')$ are the external coupling rate and the lowering operator of element E$j$ at time $t'$.
Note that in our notations the frequency and wavenumber and the time and position have the same dimensions, respectively, as we set the velocity of microwaves to be unity.
By multiplying a coefficient $e^{+ikr}/\sqrt{2\pi}$ and integrating Eq.~(\ref{rak}) from $-\infty$ to $\infty$ with respect to $k$, we obtain the input--output relation for the right-propagating mode in the real-space representation as
\eq{
\label{ior}
\vec{b}_r &=  \vec{b}_{r-t+t_{\rm i}}(t_{\rm i}) \notag \\
&-i\sum_{j=1,2}\sqrt{\frac{\gamma_j(t-r+r_j)}{2}}\,\Theta_{r\in(r_j,r_j+t)}\:\hat{L}_j(t-r+r_j),
}
where $\Theta$ is a product of Heaviside step functions, $\Theta_{r\in(a,b)} = \theta(r-a)\theta(b-r)$, which describes the time causality.
Note that we use $\int_{-\infty}^\infty dk\:e^{-ikr}=2\pi\delta(r)$.
The input--output relation for the left-propagating mode is also obtained in a similar way, i.e.\ 
\eq{
\label{iol}
\cev{b}_r &=  \cev{b}_{r+t-t_{\rm i}}(t_{\rm i}) \notag \\
&-i\sum_{j=1,2}\sqrt{\frac{\gamma_j(t+r-r_j)}{2}}\,\Theta_{r\in(r_j-t,r_j)}\:\hat{L}_j(t+r-r_j).
}

By substituting Eqs.~(\ref{ior}) and (\ref{iol}) into Eq.~(\ref{dOdt}), we obtain the derivative equation of the operator $\hat{O}$ as 
\begin{widetext}
\eq{
\label{dOdt2}
\frac{d\hat{O}}{dt} 
&= i \left[\hat{H}_{\rm s},\hat{O}\right]+i\sum_{j=1,2} \sqrt{\frac{\gamma_j}{2}}\bigg(\left[\hat{L}_j^\dag ,\hat{O}\right]\bigg(\vec{b}_{r_j-t+t_{\rm i}}(t_{\rm i}) +\cev{b}_{r_j+t-t_{\rm i}}(t_{\rm i})\bigg)+\left(\vec{b}_{r_j-t+t_{\rm i}}^{\: \dag}(t_{\rm i})+\cev{b}_{r_j+t-t_{\rm i}}^{\: \dag}(t_{\rm i})\right)\left[\hat{L}_j,\hat{O}\right]\bigg)\notag \\
&+\sum_{j=1,2}\frac{\gamma_j}{2}\left(\left[\hat{L}_j^\dag ,\hat{O}\right]\hat{L}_j-\hat{L}_j^\dag \left[\hat{L}_j,\hat{O}\right]\right) +\sum_{j=1,2}\frac{\sqrt{\gamma_j\gamma_{\bar{j}}(t-d)}}{2}\Big(\left[\hat{L}_j^\dag ,\hat{O}\right]\hat{L}_{\bar{j}}(t-d)-\hat{L}_{\bar{j}}^\dag(t-d)\left[\hat{L}_j,\hat{O}\right]\Big),
}
\end{widetext}
where $\bar{j} = 2\:(\bar{j} =1)$ if $j=1 \:(j=2)$.

Here, we employ the free-evolution approximation, $\hat{L}_j(t-d)\approx\hat{L}_j(t)\:e^{+i\omega_jd}$. 
This approximation is valid when the delay time $d$ is much shorter than the time scale of the evolution of the artificial molecule.
We also assume that the delay time $d$ is also much shorter than the time scale of the variation of the radiative decay rate, resulting in the slowly-varying-decay-rate approximation, $\gamma_j(t-d)\approx\gamma_j(t)$.
Furthermore, the initial state of the propagating mode is assumed to be the vacuum state, enabling us to neglect the annihilation operator at the initial time $t=t_{\rm i}$, i.e.\ $\vec{b}_{r_j-t+t_{\rm i}}(t_{\rm i}) \rightarrow 0$ and $\cev{b}_{r_j+t-t_{\rm i}}(t_{\rm i}) \rightarrow 0$.
Then, Eq.~(\ref{dOdt2}) is rewritten as 
\eq{
\label{dOdt3}
\frac{d\hat{O}}{dt} 
&=  i \left[\hat{H}_{\rm s},\hat{O}\right]+\sum_{j=1,2}\frac{\gamma_j}{2}\left(\left[\hat{L}_j^\dag ,\hat{O}\right]\hat{L}_j-\hat{L}_j^\dag \left[\hat{L}_j,\hat{O}\right]\right)\notag \\
+&\sum_{j=1,2}\frac{\sqrt{\gamma_1\gamma_2}}{2}\left(\left[\hat{L}_j^\dag ,\hat{O}\right]\hat{L}_{\bar{j}}\:e^{+i\omega_{\bar{j}}d}-\hat{L}_{\bar{j}}^\dag \left[\hat{L}_j,\hat{O}\right]e^{-i\omega_{\bar{j}}d}\right).
}

Using the cyclic invariance of the trace, we can express the time-derivative of the expectation value $\langle\hat{O}\rangle$ in the Schr\"{o}dinger picture, i.e.\
\begin{widetext}
\eq{
\frac{d\langle\hat{O}\rangle}{dt} = \:& {\rm Tr}\left[\frac{d\hat{O}}{dt}\:\hat{\rho}_{\rm tot}\right] ={\rm Tr}\left[\hat{O}\:\frac{d\hat{\rho}_{\rm tot}}{dt}\right].\notag \\
= &-i\:{\rm Tr}\left[\hat{O}\:[\hat{H}_{\rm s},\:\hat{\rho}_{\rm tot}]\right] +\sum_{j=1,2}\:{\rm Tr}\bigg[\hat{O}\:\gamma_j\bigg(\hat{L}_j\hat{\rho}_{\rm tot}\hat{L}_j^\dag \notag -\frac{1}{2}\left\{\hat{L}_j^\dag\hat{L}_j,\:\hat{\rho}_{\rm tot}\right\}_+\bigg)\bigg]\notag\\
&+\sum_{j=1,2}\:{\rm Tr}\bigg[\hat{O}\:\frac{\sqrt{\gamma_1\gamma_2}}{2}\bigg(\hat{L}_{\bar{j}}\hat{\rho}_{\rm tot}\hat{L}_j^\dag e^{+i\omega_{\bar{j}}d}+\hat{L}_j\hat{\rho}_{\rm tot}\hat{L}_{\bar{j}}^\dag e^{-i\omega_{\bar{j}}d} - \hat{L}_j^\dag\hat{L}_{\bar{j}}\hat{\rho}_{\rm tot}e^{+i\omega_{\bar{j}}d}- \hat{\rho}_{\rm tot}\hat{L}_{\bar{j}}^\dag\hat{L}_j e^{-i\omega_{\bar{j}}d}\bigg)\bigg],
\label{dOdtHS}
}
\end{widetext}
where $\{\cdot,\cdot\}_+$ is the anti-commutation relation.

The expectation value of the local operator $\hat{O}$ acting on the artificial-molecule subspace can be obtained independently of the partial trace of the propagating modes.
Thus, the reduced master equation for the artificial molecule is derived from Eq.~(\ref{dOdtHS}), i.e.\ 
\eq{
\frac{d\hat{\rho}}{dt}
&=  -i\left[\hat{H}_{\rm s},\:\hat{\rho}\right]+\sum_{j=1,2}\gamma_jD(\hat{L}_j)\hat{\rho} \notag\\
&+\sum_{j=1,2}\frac{\sqrt{\gamma_1\gamma_2}}{2}\Big(\hat{L}_{\bar{j}}\hat{\rho}\hat{L}_j^\dag e^{+i\omega_{\bar{j}}d} \hat{L}_j\hat{\rho}\hat{L}_{\bar{j}}^\dag e^{-i\omega_{\bar{j}}d} \notag\\
&\hspace{2cm}- \hat{L}_j^\dag\hat{L}_{\bar{j}}\hat{\rho}\:e^{+i\omega_{\bar{j}}d}- \hat{\rho}\hat{L}_{\bar{j}}^\dag\hat{L}_j e^{-i\omega_{\bar{j}}d}\Big),
}
where $\rho = {\rm Tr}_{\rm w}[\rho_{\rm tot}]$ is the density-matrix operator of the artificial molecule and ${\rm Tr}_{\rm w}[\cdot]$ is the partial trace with respect to the propagating modes.
Note that we define a superoperator for the individual decay terms as
$D(\hat{A})\hat{\rho}=\hat{A}\hat{\rho}\hat{A}^\dag-\{\hat{A}^\dag\hat{A},\hat{\rho}\}_+/2$.
Furthermore, the master equation can be rewritten in the Lindblad form as
\begin{equation}
\label{drhodt}
\begin{aligned}
\frac{d\hat{\rho}}{dt} = -i\left[\hat{H}_{\rm s}+\hat{H}_J,\:\hat{\rho}\right]
+\sum_{j,k=1,2}\gamma_{jk}D(\hat{L}_j,\hat{L}_k)\hat{\rho},
\end{aligned}
\end{equation}
where
$
\hat{H}_J = J\hat{L}_1^\dag\hat{L}_2 + J^* \hat{L}_2^\dag \hat{L}_1
$
is the waveguide-mediated energy-exchange interaction between the two elements with the strength
$
J= \frac{\sqrt{\gamma_1\gamma_2}}{2}\frac{e^{+i\omega_2d}-e^{-i\omega_1d}}{2i}
$,
$\gamma_{jj} = \gamma_j$ is the individual radiative decay rate, and 
$
\gamma_{12} = \gamma_{21}^* = \sqrt{\gamma_1\gamma_2}\:\frac{e^{+i\omega_2d}+e^{-i\omega_1d}}{2}
$
is the correlated radiative decay rate.
Here, we define a superoperator describing the correlated decay terms as
$D(\hat{A},\hat{B})\hat{\rho}=\hat{B}\hat{\rho}\hat{A}^\dag-\{\hat{A}^\dag\hat{B},\hat{\rho}\}_+/2$.

\section{Selective directional emission}
\label{appendixC}
For directional emission of a single photon, both elements should have the same frequency as the photon frequency $\omega_{\rm p}$, i.e.~$\omega_1=\omega_2=\omega_{\rm p}$.
The external coupling rates of the elements are set to be the same, i.e.\ $\gamma_1=\gamma_2=\gamma$.
In the case of using the transfer resonators, the tunable couplings between the resonators and qubits should be identical, i.e.~$g_1=g_2=g$.
Moreover, the distance between the two elements should be a quarter wavelength at the photon frequency, i.e.~$\omega_{\rm p}d = \pi/2$.
As a result, the correlated decay rate vanishes as $\gamma_{12}=\gamma_{21}=0$, while the waveguide-mediated energy-exchange interaction is maximized as
\begin{equation}
\label{HJ}
\hat{H}_J = \frac{\gamma}{2}(\hat{L}_1^\dag\hat{L}_2 + \hat{L}_2^\dag \hat{L}_1).
\end{equation} 
As discussed in the main text, a direct inter-element interaction is needed to cancel the waveguide-mediated interaction. 
In the simulation, we add an additional energy-exchange interaction, defined as 
\begin{equation}
\label{Hc}
\hat{H}_{\rm c} = g_{\rm c}(\hat{L}_1^\dag\hat{L}_2 + \hat{L}_2^\dag \hat{L}_1).
\end{equation} 
For the perfect cancellation, the coupling strength should fulfill $g_{\rm c}=-\gamma/2$.
From these conditions, the master equation~(\ref{dOdtHS}) can be rewritten as
\begin{equation}
\label{drhodt2}
\begin{aligned}
\frac{d\hat{\rho}}{dt}= -i\left[\hat{H}_{\rm s}+\hat{H}_J+\hat{H}_{\rm c},\:\hat{\rho}\right]
+\gamma\sum_{j=1,2}D(\hat{L}_j)\hat{\rho},
\end{aligned}
\end{equation}
which enables us to calculate the dynamics of the artificial molecule.

To verify the directionality emission, it is necessary to characterize the outgoing fields from the artificial molecule.
Using the density-matrix operator of the system evolving in time under Eq.~(\ref{drhodt2}) and the input--output relations of Eqs.~(\ref{ior}) and (\ref{iol}), we obtain the complex amplitude and photon flux emitted in each direction.
From Eq.~(\ref{ior}), the right-propagating output field at position $r\:(>r_2\geq0)$ is given by
\eq{
\vec{b}_r(t) &=  \vec{b}_{r-t+t_{\rm i}}(t_{\rm i}) -i\sqrt{\frac{\gamma(t-r)}{2}}\:\hat{L}_1(t-r)e^{+i\pi/4}\notag\\
&-i\sqrt{\frac{\gamma(t-r)}{2}}\:\hat{L}_2(t-r)e^{-i\pi/4},
}
where we also employ the free-evolution approximation and the slowly-varying-decay-rate approximation. 
By defining the input and output fields as $\vec{b}_{\rm out}(t) = \vec{b}_r(t+r)$ and $\vec{b}_{\rm in}(t) = \vec{b}_{-t+t_{\rm i}}(t_{\rm i})$, respectively, the input--output relation is given by
\begin{equation}
\label{boutr}
\vec{b}_{\rm out}(t) =  \vec{b}_{\rm in}(t) -i\sqrt{\frac{\gamma}{2}}\:\hat{L}_1(t)e^{+i\pi/4}-i\sqrt{\frac{\gamma}{2}}\:\hat{L}_2(t) e^{-i\pi/4}.
\end{equation}
Thus, the expectation values of the complex amplitude $\beta_{\rm R}$ and the photon flux $n_{\rm R}$ emitted rightwards at time $t$ are obtained as
\begin{equation}
\label{car}
\beta_{\rm R}(t) = \left\langle\vec{b}_{\rm out}(t)\right\rangle =  -i\sqrt{\frac{\gamma}{2}}\:\left(\left\langle\hat{L}_1\right\rangle e^{i\pi/4}+\left\langle\hat{L}_2\right\rangle e^{-i\pi/4}\right)
\end{equation}
and
\eq{
\label{pfr}
n_{\rm R}(t)&=\left\langle\vec{b}_{\rm out}^{\: \dag}(t)\vec{b}_{\rm out}(t)\right\rangle = \frac{\gamma}{2}\Big(\left\langle\hat{L}_1^\dag\hat{L}_1 \right\rangle + \left\langle \hat{L}_2^\dag\hat{L}_2 \right\rangle \notag\\
&-i\left\langle \hat{L}_1^\dag\hat{L}_2 \right\rangle +i\left\langle \hat{L}_2^\dag\hat{L}_1 \right\rangle \Big),
}
where $\langle \cdot \rangle$ on the right-hand side is the expectation value which can be obtained by the density-matrix operator of the artificial molecule at time $t$.
In the same way, the input--output relation for the left-propagating mode is given by
\begin{equation}
\label{boutl}
\cev{b}_{\rm out}(t) =  \cev{b}_{\rm in}(t) -i\sqrt{\frac{\gamma}{2}}\:\hat{L}_1(t)e^{-i\pi/4}-i\sqrt{\frac{\gamma}{2}}\:\hat{L}_2(t) e^{+i\pi/4},
\end{equation}
where the input and output modes are defined as $\cev{b}_{\rm out}(t) = \cev{b}_r(t-r)$ and $\cev{b}_{\rm in}(t)=\cev{b}_{t-t_{\rm i}}(t_{\rm i})$, respectively.
The complex amplitude $\beta_{\rm L}$ and photon flux $n_{\rm L}$ emitted leftward at time $t$ are obtained as
\begin{equation}
\label{cal}
\beta_{\rm L}(t) = \left\langle\vec{b}_{\rm out}(t)\right\rangle = -i\sqrt{\frac{\gamma}{2}}\:\left(\left\langle\hat{L}_1\right\rangle e^{-i\pi/4}+\left\langle\hat{L}_2\right\rangle e^{i\pi/4}\right)
\end{equation}
and
\eq{
\label{pfl}
n_{\rm L}(t)&=\left\langle\cev{b}_{\rm out}^{\: \dag}(t)\cev{b}_{\rm out}(t)\right\rangle = \frac{\gamma}{2}\Big(\left\langle\hat{L}_1^\dag\hat{L}_1 \right\rangle + \left\langle \hat{L}_2^\dag\hat{L}_2 \right\rangle \notag\\
&+i\left\langle \hat{L}_1^\dag\hat{L}_2 \right\rangle -i\left\langle \hat{L}_2^\dag\hat{L}_1 \right\rangle \Big).
}
In addition, the single-photon occupancy emitted in each direction is calculated by integrating the photon flux from the initial time $t=t_{\rm i}$ to the final time $t_{\rm f}$, i.e.\ 
\begin{equation}
P_{\rm R} = \int_{t_{\rm i}}^{t_{\rm f}} dt\:n_{\rm R}(t)
\quad{\rm and}\quad
P_{\rm L}= \int_{t_{\rm i}}^{t_{\rm f}} dt\:n_{\rm L}(t).
\end{equation}

\section{Directional absorption and transmission}
\label{appendixD}
As discussed in the main text, an artificial molecule for absorption and transmission should satisfy the same condition as for emission.

It is not straightforward to numerically implement single-photon inputs in a Lindblad master equation since a single-photon state in a propagating mode is strongly correlated in space.
In order to simulate the single-photon absorption and transmission in our system, we couple an ancillary qubit to the waveguide.
The qubit frequency is set to be that of the input single photon $\omega_{\rm p}$.
The ancillary qubit initialized in the excited state acts as a single-photon source to the waveguide.
Without loss of generality, a single photon is considered to be transferred from the left-hand side to our system.
In other words, the ancillary qubit is coupled only to the right-propagating mode at position $r_{\rm a}$~($<r_1\leq0$) with time-varying external coupling rate $\gamma_{\rm a}$.
Thus, the input--output relation of Eq.~(\ref{ior}) is modified as 
\eq{
\label{iore}
\vec{b}_r &= \vec{b}_{r-t+t_{\rm i}}(t_{\rm i}) \notag\\
&-i\sqrt{\gamma_{\rm a}\!\left(t-r+r_{\rm a}\right)}\,\Theta_{r\in(r_{\rm a},r_{\rm a}+t)}\:\hat{\sigma}_{\rm a}\!\left(t-r+r_{\rm a}\right)\notag \\
&-i\sum_{j=1,2}\sqrt{\frac{\gamma\!\left(t-r+r_j\right)}{2}}\,\Theta_{r\in(r_j,r_j+t)}\:\hat{L}_j\!\left(t-r+r_j\right),
}
where $\hat{\sigma}_{\rm a}$ is the lowering operator of the ancillary qubit.
The Heisenberg equation of an arbitrary operator $\hat{O}$ supported by the system subspace is given by
\eq{
\label{dOdte}
\frac{d\hat{O}}{dt} &=  i \left[\hat{H}_{\rm s},\hat{O}\right]
+\gamma_{\rm a}\left(\left[\hat{\sigma}_{\rm a}^\dag ,\hat{O}\right]\vec{b}_{r_{\rm a}}+\vec{b}_{r_{\rm a}}^{\: \dag}\left[\hat{\sigma}_{\rm a},\hat{O}\right]\right)\notag \\
&+\sum_{j=1,2} \sqrt{\frac{\gamma}{2}}\Big(\left[\hat{L}_j^\dag ,\hat{O}\right]\vec{b}_{r_j} \notag \\ 
&+\vec{b}_{r_j}^{\: \dag}\left[\hat{L}_j,\hat{O}\right]
+\left[\hat{L}_j^\dag ,\hat{O}\right]\cev{b}_{r_j}+\cev{b}_{r_j}^{\: \dag}\left[\hat{L}_j,\hat{O}\right]\Big).
}
Note that the ancillary-qubit Hamiltonian, $\omega_{\rm p}\hat{\sigma}_{\rm a}^\dag\hat{\sigma}_{\rm a}$, is added to the system Hamiltonian, $\hat{H}_{\rm s}$.
By substituting Eqs.~(\ref{iol}) and (\ref{iore}) to the Heisenberg equation (\ref{dOdte}) and employing the free-evolution approximation and the slowly-varying-decay-rate approximation,
we obtain 
\begin{widetext}
\eq{
\label{dOdt3e}
\frac{d\hat{O}}{dt} 
&= i \left[\hat{H}_{\rm s},\hat{O}\right]
+\sum_{j=1,2}\frac{\gamma}{2}\left(\left[\hat{L}_j^\dag ,\hat{O}\right]\hat{L}_j-\hat{L}_j^\dag \left[\hat{L}_j,\hat{O}\right] + \left[\hat{L}_j^\dag ,\hat{O}\right]\hat{L}_{\bar{j}}\:e^{+i\omega_{\rm p}d}-\hat{L}_{\bar{j}}^\dag \left[\hat{L}_j,\hat{O}\right]e^{-i\omega_{\rm p}d}\right)\notag \\
&+\gamma_{\rm a}\left(\left[\hat{\sigma}_{\rm a}^\dag ,\hat{O}\right]\hat{\sigma}_{\rm a}-\hat{\sigma}_{\rm a}^\dag \left[\hat{\sigma}_{\rm a},\hat{O}\right]\right)
+\sum_{j=1,2}\sqrt{\frac{\gamma\gamma_{\rm a}(t+r_{\rm a})}{2}}\left(\left[\hat{L}_j^\dag ,\hat{O}\right]\hat{\sigma}_{\rm a}\!(t+r_{\rm a})\:e^{+i\omega_{\rm p}r_j}-\hat{\sigma}_{\rm a}^\dag\!(t+r_{\rm a}) \left[\hat{L}_j,\hat{O}\right]e^{-i\omega_{\rm p}r_j}\right).
}
\end{widetext}

Considering that the ancillary qubit is located on the left-hand side and is coupled only to the right-propagating mode, it is not affected by the artificial molecule, allowing us to formally set the time delay $r_{\rm a}$ to zero. 
In a similar way as the derivation of Eq.~(\ref{drhodt2}), the reduced master equation of the composite system of the artificial molecule and the ancillary qubit is given by
\eq{
\label{drhodt2e}
\frac{d\hat{\rho}}{dt} &= -i\left[\hat{H}_{\rm s}+\hat{H}_J+\hat{H}_{\rm c}+\hat{H}_J',\:\hat{\rho}\right]\notag \\
+&\ \gamma\sum_{j=1,2}D(\hat{L}_j)\hat{\rho}+\gamma_{\rm a}D(\hat{\sigma}_{\rm a})\hat{\rho}\notag \\
+&\sum_{j=1,2}\sqrt{\frac{\gamma\gamma_{\rm a}}{2}}\left[e^{+i\omega_{\rm p}r_j} D(\hat{L}_j,\hat{\sigma}_{\rm a})\hat{\rho} + e^{-i\omega_{\rm p}r_j} D(\hat{\sigma}_{\rm a},\hat{L}_j)\hat{\rho}\right],
}
where the effective interaction Hamiltonian between the artificial molecule and the ancillary qubit is given by
$
\hat{H}_J' = \sum_{j=1,2}(J_j'\:\hat{\sigma}_{\rm a}^\dag\hat{L}_j + J_j^{'*}\:\hat{L}_j^\dag\hat{\sigma}_{\rm a})
$
with the coupling strength $J_j'=i\sqrt{\gamma_j\gamma_{\rm a}}\:e^{-i\omega_{\rm p}r_j}/2$.

In the same way as in the Appendix~\ref{appendixC}, the input--output relation of Eq.~(\ref{iore}) can be rewritten as
\eq{
\label{aore}
\vec{b}_{\rm out}(t) &=  \vec{b}_{\rm in}(t) - i\sqrt{\gamma_{\rm a}}\hat{\sigma}_{\rm a}(t) \notag \\
&-i\sqrt{\frac{\gamma}{2}}\:\left(\hat{L}_1(t) e^{+i\pi/4}+\hat{L}_2(t) e^{-i\pi/4}\right).
}
Thus, the complex amplitude and photon flux transmitted across the artificial molecule are obtained as
\eq{
\label{care}
\beta_{\rm R}(t) &= \left\langle\vec{b}_{\rm out}(t)\right\rangle \notag \\
&=  -i\sqrt{\gamma_{\rm a}}\:\langle\hat{\sigma}_{\rm a} \rangle-i\sqrt{\frac{\gamma}{2}}\:\left(\left\langle\hat{L}_1\right\rangle e^{i\pi/4}+\left\langle\hat{L}_2\right\rangle e^{-i\pi/4}\right)
}
and
\eq{
\label{pfre}
n_{\rm R}(t)&=\left\langle\vec{b}_{\rm out}^{\: \dag}(t)\vec{b}_{\rm out}(t)\right\rangle \notag \\
=\ & \gamma_{\rm a}\left\langle \hat{\sigma}_{\rm a}^\dag\hat{\sigma}_{\rm a} \right\rangle \notag\\
+& \frac{\gamma}{2}\left(\left\langle \hat{L}_1^\dag\hat{L}_1 \right\rangle + \left\langle \hat{L}_2^\dag\hat{L}_2 \right\rangle-i\left\langle \hat{L}_1^\dag\hat{L}_2 \right\rangle +i\left\langle \hat{L}_2^\dag\hat{L}_1 \right\rangle \right)\notag \\
+&\sum_{j=1,2}\sqrt{\frac{\gamma\gamma_{\rm a}}{2}}\left(\left\langle \hat{\sigma}_{\rm a}^\dag\hat{L}_j \right\rangle e^{-i\omega_{\rm p}r_j} +\left\langle \hat{L}_j^\dag\hat{\sigma}_{\rm a} \right\rangle e^{+i\omega_{\rm p}r_j} \right),
}
respectively.
In addition, the photon flux reflected by the artificial molecule is obtained using the same equation as Eq.~(\ref{pfl}).

\section{Pulse fidelity}
\label{appendixE}
\begin{figure*}[]
\begin{center}
\includegraphics{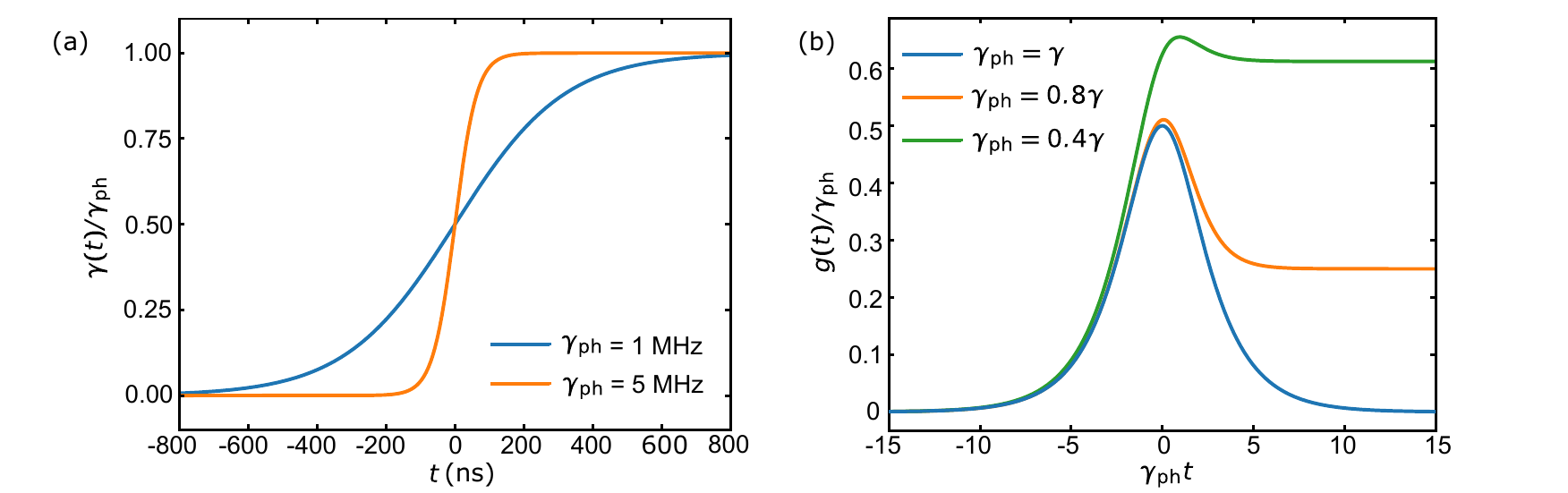} 
\caption{
Decay and coupling modulations for the emission of a photon with envelope $\phi(t)=\frac{1}{2}\sqrt{\gamma_{\rm ph}} \:\text{sech}(\gamma_{\rm ph}t/2)$. (a) Temporal modulation of the external coupling rate $\gamma(t)$ for the artificial molecule without transfer resonators. (b) Temporal modulation of the qubit--resonator coupling strength $g(t)$ in the case of the artificial molecule including the two transfer resonators. }
\label{gamma_modulation}
\end{center}
\end{figure*}
Here we describe a general framework for quantifying the pulse fidelity of the outgoing single photon within the emission and transmission protocols of the main text. 
To evaluate the fidelity of the wave packets of the single photons emitted rightwards and leftwards, as discussed in Appendix~\ref{appendixC}, the two qubits in the artificial molecule are prepared in an equal superposition of their ground states $|{\rm gg}\rangle$ and the entangled state of $|\psi_\pm\rangle =(1/\sqrt{2}) \left(  |\rm {eg}\rangle +e^{\pm i\pi/2}|\rm {ge}\rangle \right)$.
On the other hand, to characterize the fidelity of the single photon transmitted across the artificial molecule, as discussed in Appendix~\ref{appendixD}, the ancillary qubit is initialized in an equal superposition of its ground and excited states.

By using the complex amplitude of the output right- and left-propagating modes with the appropriate initial superposition state, which are calculated using Eqs.~(\ref{car}), (\ref{cal}), and (\ref{care}), the normalized wave packets of the single photons are obtained as
\begin{equation}
\label{outgoing}
\phi_{\rm R}(t) = \frac{\beta_{\rm R}(t)}{\sqrt{\int_{t_{\rm i}}^{t_{\rm f}}dt\:|\beta_{\rm R}(t)|^2}},\ 
\phi_{\rm L}(t) = \frac{\beta_{\rm L}(t)}{\sqrt{\int_{t_{\rm i}}^{t_{\rm f}}dt\:|\beta_{\rm L}(t)|^2}}.
\end{equation}
If we suppose the target wave packet of a single photon in the right- and left-propagating modes to be $\tilde{\phi}_{\rm R}(t)$ and $\tilde{\phi}_{\rm L}(t)$, the fidelities of the obtained wave packet with respect to the target are given by
\begin{equation}
\label{fidelity_in_time}
F_{\rm R} =  \left|\int_{t_{\rm i}}^{t_{\rm f}} dt\:\tilde{\phi}_{\rm R}^*(t)\phi_{\rm R}(t)\right|^2,\ 
F_{\rm L} = \left|\int_{t_{\rm i}}^{t_{\rm f}} dr\:\tilde{\phi}_{\rm L}^*(t)\phi_{\rm L}(t)\right|^2.
\end{equation}

Let us now apply this formalism to the specific case of single-photon transmitting in Fig.~\ref{figure4}(b) of the main text. Here, the expected outgoing wave packet $\tilde{\phi}_{\rm R}(t)$ in Eq.~(\ref{fidelity_in_time}) corresponds to the prepared input wave packet $\phi_{\rm R}^{\rm IN}(t)$, which is to be compared to the actual transmitted wave packet $\phi_{\rm R}(t)$, calculated from Eq.~(\ref{outgoing}).
For the transmission protocol without employing the transfer resonators, it is trivial that a single photon is faithfully transmitted when~$\gamma(t)=0$, since the two qubits in this case remain uncoupled to the waveguide.
The interesting case here is when the artificial molecule includes the two transfer resonators which decay to the waveguide at a rate $\gamma$. As explained in the main text, although the photon is also completely transmitted when the resonator--qubit coupling is switched off~[$g(t)=0$], its shape becomes distorted as a result of its interaction with the resonators. Hence, we consider here the pulse fidelity of a single photon transferring across the two resonators, in order to quantify the amount of distortion.

Before computing this fidelity, we need to find the time-delay accumulated by the transmitting photon and use it to translate back in time the transmitted wave packet, so that the shapes of the input and transmitted wave packets can be compared. 
When the two transfer resonators are degenerate at the photon frequency $\omega_{\rm p}$ owing to the perfect coupling cancellation ($\hat{H}_J+\hat{H}_{\rm c}=0$), the artificial molecule Hamiltonian without the qubit terms can be diagonalized by different orthogonal basis of the resonators, i.e.\ 
\begin{equation}
\hat{H}_{\rm s} = \sum_{\mu={\rm R,L}} \omega_{\rm p}\hat{L}_\mu^\dag\hat{L}_\mu,
\end{equation}
where $\hat{L}_{\rm R/L}=(\hat{L}_1e^{\pm i\pi/4}+\hat{L}_2e^{\mp i\pi/4})/\sqrt{2}$. 
The input--output relations of the right- and left-propagating modes of Eqs.~(\ref{boutr}) and (\ref{boutl}) can also be described by this basis, i.e.\  
\begin{equation}
\label{boutr2}
\vec{b}_{\rm out}(t) =  \vec{b}_{\rm in}(t) -i\sqrt{\gamma}\:\hat{L}_{\rm R}(t)
\end{equation}
and
\begin{equation}
\label{boutl2}
\cev{b}_{\rm out}(t) =  \cev{b}_{\rm in}(t) -i\sqrt{\gamma}\:\hat{L}_{\rm L}(t),
\end{equation}
respectively.
Therefore, the transmission coefficients from right to left or from left to right in the frequency domain are separately given by
\eq{
S_{\rm RL}(\omega) = S_{\rm LR}(\omega) = \frac{ \gamma/2+i(\omega-\omega_{\rm p}) }{\gamma/2-i(\omega-\omega_{\rm p})}.
}
From this we can obtain the phase accumulation in the transmitted pulse in the narrow bandwidth limit as
\eq{
\theta(\omega) = \arctan\left( \frac{\gamma(\omega-\omega_{\rm p})}{\gamma^2/4 - ( \omega-\omega_{\rm p} )^2} \right),
}
and hence the group delay
\eq{
\left. \frac{\partial \theta(\omega)}{\partial \omega}\right|_{\omega = \omega_\mathrm{p}} = -\frac{4}{\gamma}.
}
Thus, the wave packet accumulates a time-delay of $4/\gamma$ as it gets transmitted through the two transfer resonators. To compute the fidelity in Eq.~(\ref{fidelity_in_time}), we therefore translate back in time the transmitted wave packet by $4/\gamma$. The pulse fidelity of the transmitted wave packet is thus given by
\begin{equation}
\label{fideltiy_for_transmission}
F_{\rm R} =  \left|\int_{t_{\rm i}}^{t_{\rm f}} dt\:{\phi}_{\rm R}^{{\rm IN}*}(t)\phi_{\rm R}(t+4\gamma^{-1})\right|^2.
\end{equation}

\section{Target pulse shape}
\label{appendixF}

The wave packet of an emitted single photon is controlled by dynamically tuning the emission rate of the artificial molecule.
In this paper, the target wave packet is chosen to be the following hyperbolic secant function: 
\begin{equation}
\label{target}
\phi(t) = \frac{1}{2}\sqrt{\gamma_{\rm ph}}\:{\rm sech}\!\left(\frac{\gamma_{\rm ph}t}{2}\right),
\end{equation}
where $\gamma_{\rm ph}$ is the photon bandwidth.
Importantly, our target wave packet is time-symmetric, allowing us to perfectly absorb the single photon by simply implementing the time-reversal of the modulation used for the emission of the photon.

In order to generate the wave packet using two qubits,
the tunable external coupling rate should be temporally modulated as
\begin{equation}
\gamma(t) = \gamma_{\rm ph}\frac{{\rm sech}^2\!\left(\frac{\gamma_{\rm ph}t}{2}\right)}{1-\tanh\!\left(\frac{\gamma_{\rm ph}t}{2}\right)}.
\end{equation} 
This is also the modulation used for the ancillary qubit, which plays the role of single-photon source in the simulation of the absorption and transmission protocols of the main paper. This modulation of the external coupling rate is plotted on Fig.~\ref{gamma_modulation}(a) for different values of $\gamma_{\rm ph}$.

For the artificial molecule including two transfer resonators, the tunable coupling strength between the resonator and qubit should be modulated as
\begin{equation}
g(t) = \frac{\gamma_{\rm ph}}{4\cosh\left(\frac{\gamma_{\rm ph}t}{2}\right)}\frac{1-e^{\gamma_{\rm ph}t}+(1+e^{\gamma_{\rm ph}t})\gamma/\gamma_{\rm ph}}{\sqrt{(1+e^{\gamma_{\rm ph}t})\gamma/\gamma_{\rm ph}-e^{\gamma_{\rm ph}t}}},
\end{equation}
where the single-photon bandwidth $\gamma_{\rm ph}$ should be less than that of the transfer resonators $\gamma$~\cite{kurpiers_deterministic_2018}. This tunable coupling strength is plotted on Fig.~\ref{gamma_modulation}(b) for different values of the ratio $\gamma_{\rm ph}/\gamma$.

\section{Circuit implementation}
\label{appendixG}
\begin{figure}[t]
\begin{center}
  \includegraphics[]{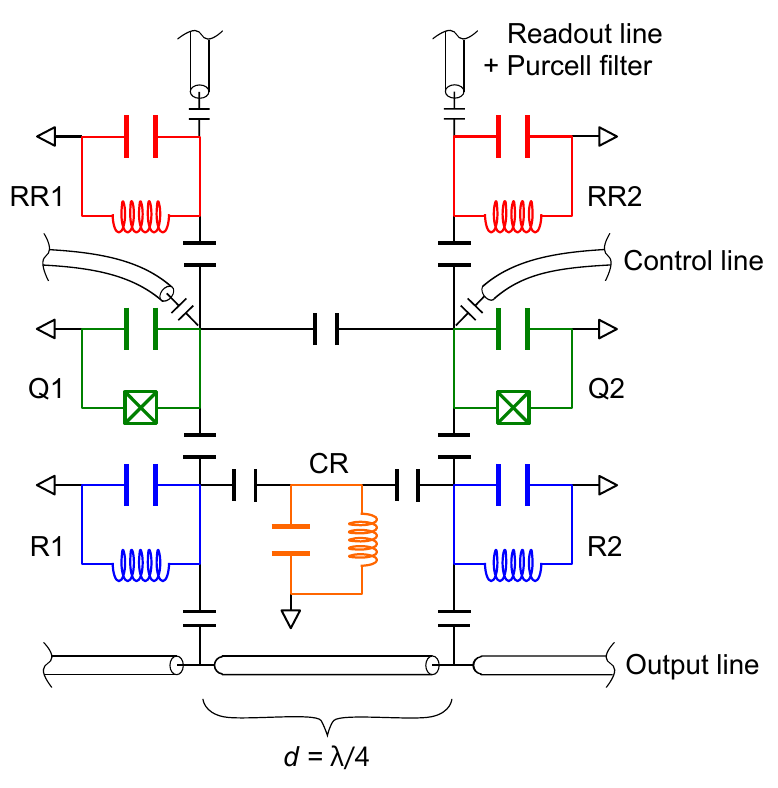} 
\caption{
Example of a possible circuit implementation of the proposed artificial molecule.
}
\label{circuit_implementation}
\end{center}
\end{figure}
Here we describe a possible circuit implementation of our second proposal shown in Fig.~~\ref{figure2}(d).
We present an illustration of this circuit in Fig.~\ref{circuit_implementation}.

Each qubit (green) is realized by a Josephson junction shunted by a large capacitor, known as a transmon qubit.
The qubits are individually controlled by a resonant microwave field applied through a weakly-coupled control line.
To create an entangled state of the two qubits, a two-qubit gate is performed on the qubits by using the capacitive coupling between them. The frequency detuning between the qubits is designed to be large enough so as to suppress the residual state-dependent frequency shift.
By driving one of the qubits with a microwave field at the resonance frequency of the other qubit, a cross-resonance gate can be realized, allowing us to naturally implement a CNOT gate between the two qubits~\cite{rigetti_fully_2010}.
Since an arbitrary two-qubit unitary operation can be realized using a combination of CNOT gates and single-qubit gates, we can prepare the two qubits in the target entangled state, $(1/\sqrt{2})(|{\rm eg}\rangle+e^{\pm\pi/2}|{\rm ge}\rangle)$.
To calibrate these operations, the states of the qubits should be read out individually.
Thus, qubit Q$j$ ($j=1,2$) is dispersively coupled to a readout LC resonator RR$j$ (red). Each resonator is connected to a readout line with a Purcell filter~\cite{jeffrey_fast_2014}.
The states of the qubits are read out by measuring readout pulses reflected by the readout resonators.

To emit a pulse-shaped microwave photon to a waveguide (output line in Fig.~\ref{circuit_implementation}), qubit Q$j$ with the resonance frequency $\omega_{{\rm Q}j}$ should be tunably coupled to transfer resonator R$j$ (blue) with the resonance frequency $\omega_{\rm R}$.
Here, we use a microwave-induced interaction between the second excited state of transmon qubit Q$j$ ($|{\rm f0}\rangle_j$) and the single photon state of transfer resonator R$j$ ($|{\rm g1}\rangle_j$)~\cite{pechal_microwave-controlled_2014}. In this scheme, in order to suppress the energy-exchange interaction in the absence of the microwave drive, it is important that qubit Q$j$ and transfer resonator R$j$ be dispersively coupled, i.e.\ $\tilde{g}_j\ll \Delta_j$, where $\tilde{g}_j$ is the Q$j$--R$j$ coupling strength and $\Delta_j=\omega_{{\rm Q}j}-\omega_{\rm R}$ is the frequency detuning.

When a microwave field is applied to the qubit at the transition frequency between the states $|{\rm f0}\rangle_j$ and $|{\rm g1}\rangle_j$, the effective Hamiltonian in the frame rotating at the drive frequency is given by
\begin{equation}
\hat{H}_{\rm s} =\sum_{j=1,2} g_j(t)\left(|{\rm f0}\rangle_{j\:j}\langle{\rm g1}| + |{\rm g1}\rangle_{j\:j}\langle{\rm f0}|\right),
\end{equation}
with the tunable coupling strength of $g_j(t)=\tilde{g}_j\alpha_j\Omega_j(t)/[\sqrt{2}\Delta_j(\Delta_j+\alpha_j)]$, where $\Omega_j(t)$ is the time-dependent qubit drive amplitude and $\alpha_j$ is the anharmonicity of transmon qubit Q$j$.
Note that we omitted in this Hamiltonian terms unrelated to this interaction. The advantages of the microwave-induced interaction are two-fold. First, the coupling strengths $g_1(t)$ and $g_2(t)$ can be accurately matched to each other by tuning the qubit drive amplitudes. Secondly, the frequency of the emitted photon can also be matched within the bandwidth of the transfer resonator by tuning the qubit drive frequency.
Note that before driving the qubits for inducing the tunable couplings, the excitations in the system are to be transferred to the upper levels by applying a $\pi$ pulse between $|{\rm e}\rangle_j$ and $|{\rm f}\rangle_j$, which brings the qubits in the new entangled state $(1/\sqrt{2})(|{\rm fg}\rangle+e^{\pm i \pi/2}|{\rm gf}\rangle)$. Once in this state, the artificial molecule is ready for directional photon emission. This scheme was implemented in a number of circuit QED experiments, an example of which is the experiment presented in Ref.~\cite{ilves_-demand_2020}.
\begin{figure*}[t]
\begin{center}
  \includegraphics[]{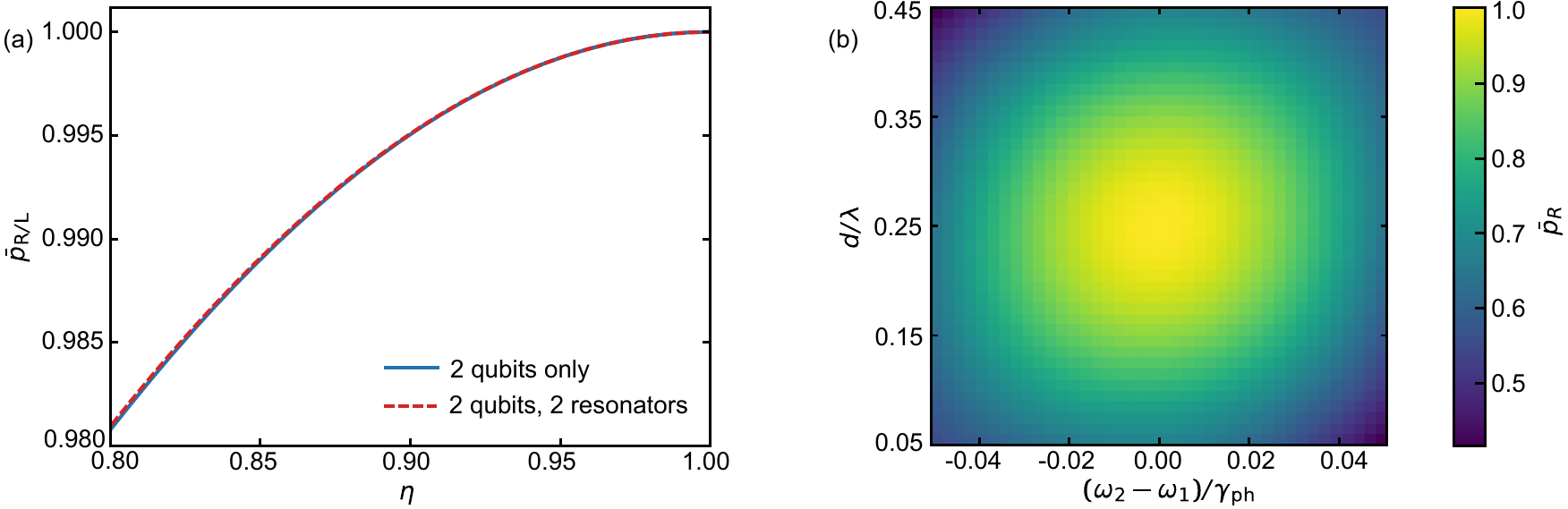} 
\caption{
Robustness of directionality. (a) Directionality $\bar{p}_{\rm R/L}$ as a function of the coupling cancellations factor $\eta$, for the designs shown in Fig.~\ref{figure2}(a) [2 qubits only] and Fig.~\ref{figure2}(d) [2 qubits, 2 resonators] of the main text. (b) Directionality $\bar{p}_{\rm R}$ as a function of the normalized qubit frequency difference and of the normalized inter-resonator distance, for the design shown in Fig.~\ref{figure4}(a) of the main text.
}
\label{directionality_vs_cancellation_ratio}
\end{center}
\end{figure*}
The transfer resonators are separately coupled with the static external coupling rate $\gamma$ to the waveguide a quarter wavelength apart. 
As described in Eq.~(\ref{HJ}), for this separation, the waveguide-mediated energy-exchange interaction between the transfer resonators is maximized.
To cancel the interaction, coupling resonator CR (orange) with the resonance frequency $\omega_{\rm C}$ is inserted between the transfer resonators.
By designing this coupling resonator to be dispersively coupled to the transfer resonators, a virtual photon in the coupling resonator induces an additional effective energy-exchange interaction between the transfer resonators, as described in Eq.~(\ref{Hc}).
In this configuration, the effective coupling strength is given by $g_{\rm c}=g_{\rm C1}g_{\rm C2}/\Delta_{\rm C}$, where $g_{{\rm C}j}$ is the coupling strength between the resonators R$j$ and CR, and $\Delta_{\rm C}= \omega_{\rm R}-\omega_{\rm C}$.
In the example of Fig.~\ref{circuit_implementation}, since the signs of the coupling strengths of $g_{{\rm C}1}$ and $g_{{\rm C}2}$ are the same, the negative coupling strength of $g_{\rm c}$ is realized by ensuring that $\Delta_{\rm C}$ is negative. 
Thus, the complete cancellation of the waveguide-mediated interaction is achieved by designing a device with $g_{\rm c}=-\gamma/2$.\\
\indent There are a few conditions that need to be met in order to achieve high fidelities of the directional emission, absorption, and transmission.
The first is that the communication time between the qubit and the photonic mode in the waveguide be much faster than the lifetime of the qubit.
This communication time is of the order of $\max [g_j (t)]\approx\gamma$.
The lifetime of the qubit is intrinsically limited by the Purcell effect induced by the coupling to the transfer resonator, where the Purcell decay rate of qubit Q$j$ is given by $\Gamma_j=(\tilde{g}_j/\Delta_j)^2\gamma$.
Therefore, this condition is achieved in the dispersive limit where $\max [g_j (t)]\approx\gamma\gg(\tilde{g}_j/\Delta_j)^2\gamma=\Gamma_j$.
The second condition is that the dispersive shift of the transfer resonator depending on the qubit state be much smaller than the frequency bandwidth of the transfer resonator.
This is required to prevent a photonic mode transmitted across a non-targeted node from being entangled with it.
The dispersive shift is given by $\chi_j=(\tilde{g}_j/\Delta_j)^2\alpha_j$.
Thus, the condition $\chi_j\ll\gamma$ can be rewritten as $(\tilde{g}_j/\Delta_j)^2\alpha_j\ll \max [g_j (t)]$, which in turn can be approximated as $\tilde{g}_j\ll\max [\Omega_j(t)]$ when $\Delta_j\gg\alpha_j$. This is a condition which can be met using properly designed superconducting circuits.

\section{Robustness of directionality against implementation imperfections}
\label{appendixH}

Here we investigate the effect of implementation imperfections on the directionality of emission $\bar{p}_{\rm R/L}$. 

One type of defect could be the imperfect implementation of the cancellation coupling. The directionality $\bar{p}_{\rm R/L}$ is defined as the ratio of the probability $P_\mathrm{R/L}$ for the photon to be emitted in the intended direction of emission to the total photon-emission probability:
\eq{
\label{directionality}
\bar{p}_{\rm R/L} = \frac{P_{\rm R/L}}{P_{\rm R} +P_{\rm L}}.
}
In order to study the effect of imperfect cancellation, we introduce a coupling cancellation factor $\eta$ in the cancellation Hamiltonian of Eq.~(\ref{Hc}), which becomes
\begin{equation}
\label{Hc2}
\hat{H'}_{\rm c} = \eta g_{\rm c}(\hat{L}_1^\dag\hat{L}_2 + \hat{L}_2^\dag \hat{L}_1).
\end{equation} 
The $\eta$ factor now allows us to tune the amount of cancellation: $\eta=1$ corresponds to perfect cancellation, while $\eta=0$ corresponds to no cancellation at all. In Fig.~\ref{directionality_vs_cancellation_ratio}(a), we plotted the directionality versus the cancellation factor, for the two-qubit system without~(blue) and with~(red) the transfer resonators, respectively. Clearly, for both systems proposed in the main text, the directionality turns out to be very robust against a residual effective coupling. In particular, for a residual coupling of 10\%, the directionality decreases only by $0.5\%$.\\
\indent Furthermore, the inter-qubit distance and qubit frequencies could differ from their ideal values. To investigate the effect of such imperfections in the case of our first proposal [Fig.~\ref{figure2}(a)], we show in Fig.~\ref{directionality_vs_cancellation_ratio}(b) a 3D plot of the directionality $\bar{p}_{\rm R}$ defined in Eq.~(\ref{directionality}) as a function of the normalized frequency difference between the two qubits $(\omega_2-\omega_1)/\gamma_{\rm ph}$, and of the normalized inter-qubit distance $d/\lambda$. Clearly, the directionality is relatively insensitive to deviations of the inter-qubit distance from $d/\lambda=0.25$, with a deviation of $0.02\lambda$ from $d=0.25\lambda$ leading to a decrease in directionality by less than $1\%$. On the other hand, obtaining higher than $99\%$ directionality will require reaching qubit frequencies that do not differ by more than $|\omega_2-\omega_1|/\gamma_{\rm ph} \simeq 0.008$. For qubit frequencies aimed at $\omega_{1,2}/(2\pi) = 8$ GHz and for a photon bandwidth of $\gamma_{\rm ph}/(2\pi) = 10$ MHz, this difference corresponds to $|\omega_2-\omega_1|/(2\pi) \simeq 80$ kHz. Note that this 2D map can also be applied to our second proposal [Fig.~\ref{figure2}(d)] by simply replacing the qubit frequencies $\omega_{1,2}/(2\pi)$ by the emission frequencies of the transfer resonators. In this case, as explained in Appendix~\ref{appendixG}, it is the frequency of the parametric drives that will determine the individual emission frequencies of the resonators.

\bibliography{directional_emission_paper.bib,mybib.bib}

\end{document}